# Light Curve Modeling of Eclipsing Binary Systems with Delta Scuti Component

J. A. D. M. Dharmathilaka.[1], J. Adassuriya[2], K. P. S. C. Jayaratne [2], Jordi L. Gutiérrez[3]

[1]*Department of Physical Sciences and Technology, Faculty of Applied Sciences, Sabaragamuwa University of Sri Lanka, Beihuloya, Sri Lanka*
[2]*Astronomy and Space Science Unit, Department of Physics, Faculty of Science, University of Colombo, Sri Lanka*
[3]*Department of Physics, Universitat Politècnica de Catalunya (UPC), Spain*

janaka@phys.cmb.ac.lk

## 1. ABSTRACT

Astroseismology in eclipsing binaries with Delta (δ) Scuti components offers a powerful means to derive stellar parameters and probe internal structures. To enable accurate frequency analysis, binary characteristics must be disentangled from the observed light curves. This study utilizes the LC2015 light curve modeling method, followed by the DC2015 differential correction process, integrated into the Wilson-Devinney (WD) eclipsing binary modeling code. The analysis focuses on two δ Scuti binary systems, KIC 8504570 and SX DRACONIS (Dra), using Kepler and TESS photometric data, supplemented by literature-derived initial stellar parameters. The DC2015 process employs the Levenberg-Marquardt algorithm to minimize the difference between observed and modeled light curves. The refined models provide highly accurate stellar parameters, including primary and secondary star temperatures ($T_{eff,1}$ and $T_{eff,2}$), mass ratio (q), and primary star luminosity ($L_1$) with associated errors. For KIC 8504570: $T_{eff,1}$ (7400.9 ± 1.6) K, $T_{eff,2}$ (5450.4 ± 1.0) K, q (0.5208 ± 0.0002), and $L_1$ (11.8043 ± 0.0006)$L_\odot$. For SX Dra: $T_{eff,1}$ (7729.7 ± 1.1) K, $T_{eff,2}$ (4927.5 ± 0.5) K, q (0.4772 ± 0.0006), and $L_1$ (7.0474 ± 0.0015)$L_\odot$.

## 2. INTRODUCTION

Pulsating variables are the intrinsic variable stars that exhibit the change of stars' brightness due to the physical changes of the star itself. It can happen periodically by the contraction and the expansion of the star surface layers. δ Scuti type stars are one of the most important star categories under the pulsating variable stars, which are located inside the classical instability strip in the Hertzsprung-Russell (HR) Diagram. They pulsate with short periods as well as have radial and non-radial pulsation modes with frequencies ranging between 3 and 80 cycles per day ($d^{-1}$). Their spectral types range between AV to FV, and they have an intermediate mass range between 1.4 and 2.5 $M_\odot$.

The luminosity classes of the δ Scuti type stars typically range from dwarf to giant [1, 8, 9].

A binary star system is a combination system of two stars. They are gravitationally bound to each other, and both stars orbit around a common center of mass, exhibiting an elliptical path of orbit [8]. An eclipsing binary is a subclass of binary system consisting of a pair of stars that orbit each other in a way that the light of one star is blocked by the other star as seen from Earth. The presence of the δ Scuti type pulsating variable component in an eclipsing binary system makes it a valuable system for the analysis, particularly in the field of asteroseismology. Asteroseismology is a study of understanding the interior structures of the stellar systems [6]. This combination is extremely important because it is a combination of two different concepts in astrophysics yielding remarkable results. Therefore, studying such systems is critical. According to the studies of Lee et al. (2016b) and Soydugan et al. (2016), many systems with δ Scuti stars in eclipsing binaries have been discovered [6]. Basically, such systems can be divided in to two categories as semi-detached and detached binaries. If one star fills its Roche Lobe and transfer matter to the other star that type of systems can be categorized as semi-detached binaries while if each star remains within their own Roche Lobe that type of systems can be categorized as detached binaries. δ Scuti pulsators in semi-detached eclipsing binaries was defined as oscillating Eclipsing Algol (oEA) systems by Mkrtichian et al. (2004) [5]. Liakos & Niarchos (2015) studied how the primary δ Scuti component evolves more slowly through the main sequence compared to the single δ Scuti stars due to the mechanisms of mass transfer and the tidal distortion in such binaries [6, 8]. Soydugan et al. (2006b) first noted the connection between pulsation and the orbital periods in eclipsing binaries with δ Scuti components by considering 20 systems [6].

Pulsation modeling and the asteroseismic study of eclipsing binaries consisting with δ Scuti components are crucial and highly valuable in astronomy. To achieve accurate results, the initial steps involve light curve modeling and disentangle the binary characteristics from the light curves. This process must be carried out more precisely to ensure accurate results not only for the pulsation model, but also for the asteroseismic study. Therefore, this study mainly focuses on the precise study of light curve modeling and disentangle of binary characteristics from the light curves. For the purpose, Wilson – Devinney (WD) eclipsing binary modeling code was used [3, 7]. Two key processes included in WD code are the Light Curve modeling process (LC2015) and Differential Correction process (DC2015) which deal with the light curve modeling and stellar parameter refinement process respectively [3, 7].

The Kepler Space Telescope and the Transiting Exoplanet Survey Satellite (TESS) are two space telescopes which provide high quality photometric data to investigate phenomenon and understand the structure of binary and pulsating systems [4, 12]. TESS observes in both short cadence (SC) mode with 2 minutes and long cadence (LC) mode with 30 minutes. Kepler also has been observed in both short cadence (SC) mode with 1 minute and long cadence (LC) mode with 30 minutes. In this study, two stars were used for the analysis namely, KIC 8504570, a detached eclipsing binary with δ Scuti component and SX Dra, a semi-detached eclipsing binary with δ Scuti component.

The article is arranged as follows. Information about the photometric observations and data retrieval methods are provided in Section 3. Binary light curve modeling by pyWD2015 LC Program is discussed in Section 4 which also includes the preparation of light curves. The results of the binary light curve modeling by pyWD2015 LC program and the stellar parameter values were also included in Section 4. Section 5 covers stellar parameter refinement process by pyWD2015 DC Process. Finally, Section 6 provides a summary of this work, along with discussion and conclusions.

## 3. OBSERVATIONS
### 3.1. Photometric Data

In this study, TESS and Kepler data were mainly used for the targets. KIC 8504570 has been observed in both "short cadence" (SC) and "long cadence" (LC) modes. Long cadence data were mainly used for the analysis, which they covered long exposure times. This system has an orbital period of 4.0077046 days. The aperture photometry data of Quarter 13 were retrieved from the Mikulski Archive for Space Telescopes (MAST Archive[1]) covering approximately 85 days which includes about 21 orbital cycles.

The SX Dra system was observed by the TESS mission with ID number TIC 353854078 and has an orbital period of 5.169412 days. It has been observed in several sectors and sector 57 data were mainly used in this study. The aperture photometry data of Sector 57 was retrieved from the NASA MAST using the lightkurve package covering approximately 29 days which includes about 5 orbital cycles.

The basic stellar parameters of the two targets were obtained from A. Liakos and P. Niarchos (2020) [9] and E. Soydugan and Y. Kacar (2013) [11]. The parameters obtained in [9] and [11] were determined by only running LC2015 model. However, the differential correction method (DC) can be used to further converge the models to the observations using Levenberg-Marquardt algorithm and thereby fine-tuning the stellar parameter. To estimate the relevance of this effect and to compute a best binary model with parameters, the pyWD2015 DC modeling package was used.

The ephemeris of a variable star is a mathematical expression that predicts the times when the star will reach specific brightness levels, typically marking the moment of its deepest eclipse in the case of an eclipsing binary system [10]. The resulting ephemeris is calculated by;

$$T_{epoch} = T_0 + (P \times E) \tag{1}$$

In this equation (1), $T_0$ is the initial time of minimum light in the light curve, usually the deepest eclipse (primary eclipse), P is the orbital period of the system and E is the number of orbital cycles that have occurred from initial time to the reference time. In

---
[1] https://archive.stsci.edu/missions-and-data/kepler

pyWD2015 binary star modeling, the epoch represents the reference time of the dataset at the primary eclipse ($T_{epoch}$).

## 4. BINARY LIGHT CURVE MODELING BY PYWD2015 LC PROGRAM

KIC 8504570 and SX Dra systems have been modeled solely on the basis of pyWD2015 LC model. The stellar parameters obtained from LC2015 can be further refined using DC2015 model. Therefore, the light curves were subjected to DC model.

### 4.1. Preparation of the Light Curves

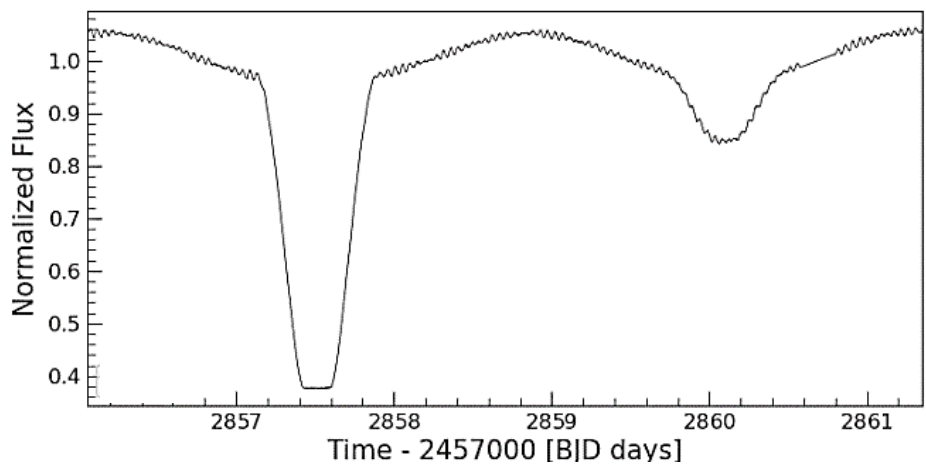

Figure 1: Blow up of a section of the light curve in SX Dra

The retrieved data for both targets were cleaned of obvious outliers. Both systems of KIC 8504570 and SX Dra were normalized to the median value of Quarter 13 and Sector 57, respectively. The light curve and the blow-up of a section of the light curve (which shows the pulsations and eclipses in greater detail) corresponding to Q13 of KIC 8504570 and S57 of SX Dra were designed in this study. Figure 1 represents the blow-up of a section of the light curve of SX Dra system. Throughout this article only represents the diagrams for the results obtained for SX Dra system.

### 4.2. The Binary Light Curve Modeling by pyWD2015 LC Program and Results

The LC analysis was done by using pyWD2015 software [3, 7]. Most of the parameter values for both systems were obtained thanks to the studies of A. Liakos and P. Niarchos (2020) [9] and E. Soydugan and Y. Kacar (2013) [11], for KIC 8504570 and SX Dra respectively. The temperature of the primary component ($T_{eff,1}$) was kept as fixed during the modeling, while temperature of the secondary component ($T_{eff,2}$) was adjusted to get the best fit for the system. Gravity darkening coefficients (g) and Albedo Coefficients (A) were assigned fixed at their theoretical values where g=1 and A= 1 for radiative envelopes and g=0.3 and A=0.5 for convective envelops [2]. Additionally, the linear limb darkening coefficients (x) were also assigned as fixed values, they were determined according to the study of Van Hamme [13]. For this modeling, the semi-major axes (a) were determined with the help of Kepler's Third Law. In the light curve modeling, the values of inclination angle (i), the modified surface potentials ($\Omega_1$, $\Omega_2$), luminosity of the primary component ($L_1$), and epoch ($T_{epoch}$) were adjusted, rather than the $T_{eff,2}$.

The system and absolute parameters resulting from the LC modeling are listed in Table 2 for KIC 8504570 and SX Dra. The LC modeling for SX Dra system is plotted in Figure 2. Lower part of the Figure 2 illustrates the LC residuals after disentangling the binary characteristics from the original light curves.

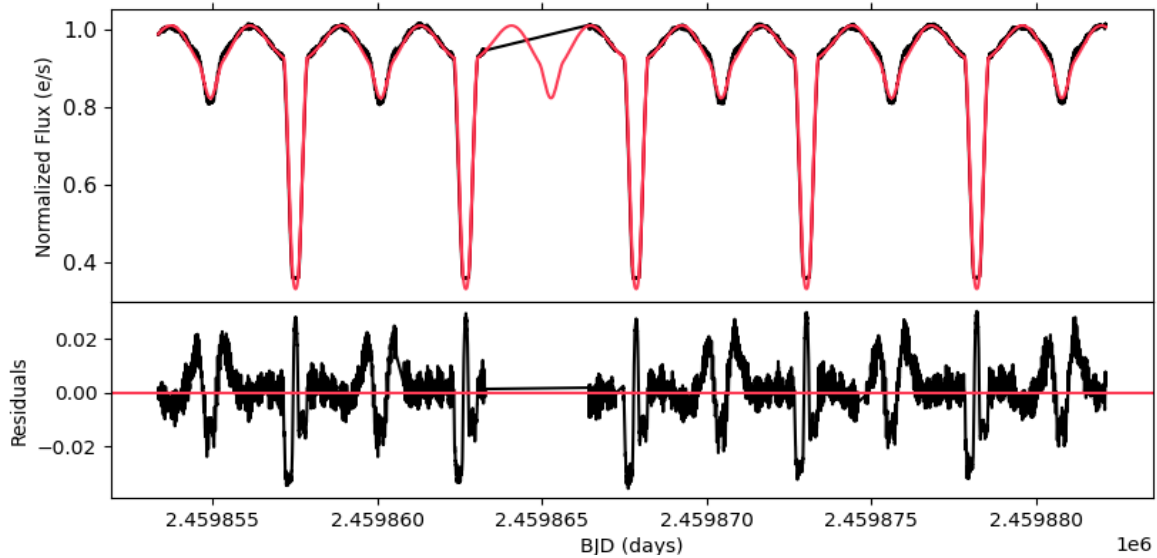

Figure 2: The results of light curve modeling of the SX Dra systems using the pyWD2015 LC program. Observed light curve (black points) for SX Dra system and theoretical LC2015 model (red points) in the upper part. The residuals of the fitting model (black points) after subtraction of the LC models represented in the lower part.

Table 1: LC2015 Modelling Results for KIC 8504570 and SX Dra Systems

| | Input Stellar System Parameters for the pyWD2015 LC Program | | | | | |
|---|---|---|---|---|---|---|
| | KIC 8504570 | SX Dra | | | KIC 8504570 | SX Dra |
| a ($R_\odot$) | 14.82* | 17.5* | | q ($M_2/M_1$) | 0.52* | 0.423* |
| i (degrees) | 84.6 | 81.3 | | $P_{orb}$ (days) | 4.0077046* | 5.169412* |

| | Input Stellar Parameters for the pyWD2015 LC Program | | | | | Output Stellar Parameters Generated by the pyWD2015 LC Program | | | |
|---|---|---|---|---|---|---|---|---|---|
| | Primary | | Secondary | | | Primary | | Secondary | |
| | KIC 8504570 | SX Dra | KIC 8504570 | SX Dra | | KIC 8504570 | SX Dra | KIC 8504570 | SX Dra |
| $T_{eff}$ (K) | 7400* | 7762* | 5450 | 4900 | M ($M_\odot$) | 1.789 | 1.891 | 0.9301 | 0.7998 |
| $\Omega$ | 7.76 | 7.568 | 9.57 | - | R ($R_\odot$) | 2.0498 | 2.4762 | 0.9254 | 5.3741 |
| A | 1* | 1* | 0.5* | 0.5* | Log g (cm s$^{-2}$) | 4.07 | 3.93 | 4.47 | 2.88 |
| g | 1* | 1* | 0.3* | 0.3* | $M_{bol}$ (mag) | 2.12 | 1.5 | 5.17 | 1.82 |
| x | 0.540* | 0.660* | 0.522* | 0.800* | Log ($L/L_\odot$) | 1.0526 | 1.2998 | -0.1695 | 1.1737 |
| L ($L_\odot$) | 11.8 | 7.31 | - | - | | | | | |

Note: * Fixed during the fitting procedure

## 5. STELLAR PARAMETER REFINEMENT PROCESS BY PYWD2015 DC PROCESS

In this section, refinement process for the stellar parameters was obtained by DC2015 process. This section discuses re-fitting of the light curve which began with the output from LC2015 and the derivation of parameter uncertainties. DC2015 is a refinement process used to further improve the model parameters after the initial fitting performed by LC2015. It uses an iterative method to fine-tune our model and minimize the deviation between the modeled light curve and the observed curve [3, 7].

### 5.1. Determination of External Iterations Number

In this DC2015 process, we need to choose a data file name from the combo box. Then, DC widget plots observational data from the selected file, together with the computed theoretical model. Here we have set external iterations number to 15. While running this, at the end of this last iteration, we noticed that the green colors appeared (which says differential correction has converged) under "output" label in "Results" tab. After convergence is achieved, the differential correction process reaches a point where the model parameters no longer change appreciably with further iterations. That means fit cannot be improved further. The best fit with best values of stellar parameters was obtained for 15 external iterations for SX Dra binary system while for KIC 8504570 binary system, the best fit was obtained for 6 external iterations.

### 5.2. Final Synthetic Light Curve and Results from DC2015 Process

After generating the optimal model for our target systems, we successfully obtained refined values for the stellar parameters through the pyWD2015 DC process. The best-fitting model for the SX Dra system, based on the observed light curve, is shown in Figure 3(a), while the phased light curve of the SX Dra system with its final fit is depicted in Figure 3(b). A similar outcome was achieved for the KIC 8504570 system.

As illustrated in Figure 2, we observed that the eclipses were not well reproduced by the model in the LC program, although this represented the best fit within the LC framework. In contrast, as shown in Figure 3(a), the discrepancies observed in the LC model, especially during the eclipses, were significantly reduced under the DC approach. Therefore, the DC model provided a superior fit for the light curves of both systems, resulting in a more accurate representation of the observed data.

Using the DC2015 process, we have refined the parameter values for inclination (i), effective temperatures ($T_{eff,1}$ and $T_{eff,2}$), surface potentials ($\Omega_1$ and $\Omega_2$), mass ratio (q), and luminosity ($L_1$). The system parameters, along with the derived absolute parameters obtained from the DC model, are presented in Table 2 for both binary star systems. These refined values provide a more accurate characterization of the systems, enhancing our understanding of their physical properties.

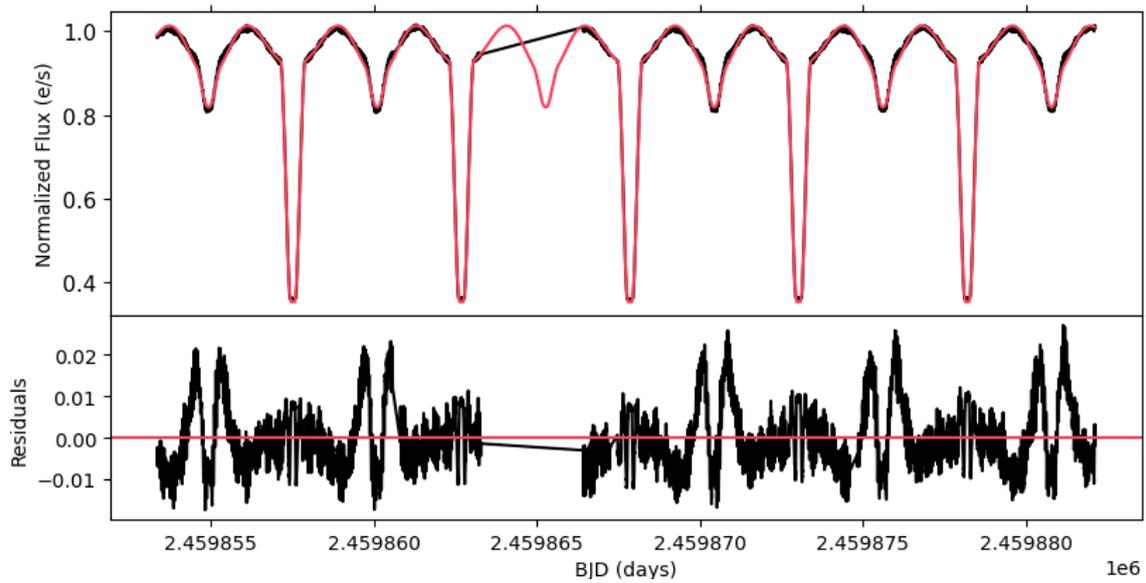

(a)

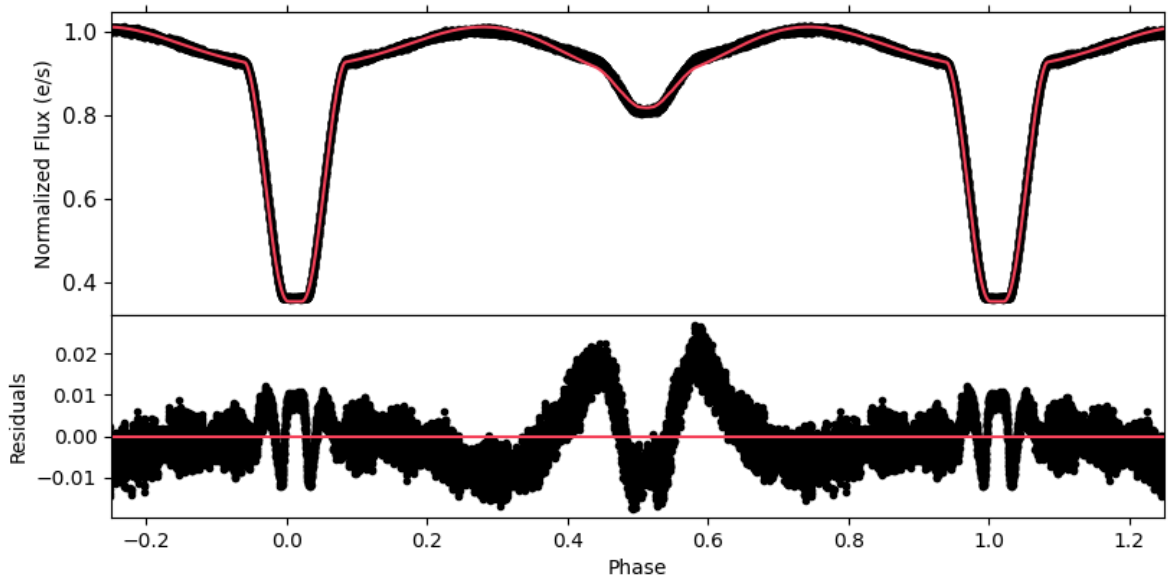

(b)

Figure 3: The results of light curve modeling of the SX Dra system using the pyWD2015 DC program. Panel (a): Observed light curve (black line) for SX Dra system and theoretical DC2015 model (red line) in upper part. Lower part represents the residuals of the best fitting model (black line) after subtraction of the DC models. Panel (b): The phased light curve of the SX Dra (black points) system with its final fit (red line) in upper part and in the lower part represents the residuals of the best fitting model (black points) after subtraction of the model.

Table 2: Resulting stellar parameters from the DC2015 model, along with their corresponding uncertainties (in parentheses)

| | Output Stellar Parameters Generated by the pyWD2015 DC Program | | | | | | | |
|---|---|---|---|---|---|---|---|---|
| | KIC 8504570 | SX Dra | | | | KIC 8504570 | SX Dra | |
| i (degrees) | 84.529 (0.006) | 81.804 (0.011) | | | $q (M_2/M_1)$ | 0.5208 (0.0002) | 0.4772 (0.0006) | |
| | Primary | | Secondary | | | Primary | | Secondary | |
| | KIC 8504570 | SX Dra | KIC 8504570 | SX Dra | | KIC 8504570 | SX Dra | KIC 8504570 | SX Dra |
| $T_{eff}$ (K) | 7400.9 (1.6) | 7729.7 (1.1) | 5450.4 (1.0) | 4927.5 (0.5) | $M (M_\odot)$ | 1.788 | 1.821 | 0.931 | 0.8695 |
| $\Omega$ | 7.7584 (0.0025) | 7.5282 (0.0018) | 9.5560 (0.0033) | - | $R (R_\odot)$ | 2.0499 | 2.4869 | 0.9281 | 5.5427 |
| $L (L_\odot)$ | 11.8043 (0.0006) | 7.0474 (0.0015) | - | - | Log g (cm s$^{-2}$) | 4.07 | 3.91 | 4.47 | 2.89 |
| | | | | | $M_{bol}$ (mag) | 2.12 | 1.51 | 5.17 | 1.72 |
| | | | | | Log ($L/L_\odot$) | 1.0527 | 1.2961 | -0.1666 | 1.2105 |

## 6. SUMMARY, DISCUSSION AND CONCLUSIONS

Under this study, a comprehensive analysis of the KIC 8504570 and SX Dra binary systems containing a δ Scut pulsating star was performed. Thanks to the Kepler and TESS missions, quality data were gathered and stellar parameter values were obtained from the literature and using theoretical concepts. The eclipsing light curves of the both systems were modeled using the WD eclipsing binary modeling code. In the pyWD2015 eclipsing binary modeling, LC program and DC program were applied to both star systems. The output from LC2015 program includes a set of stellar parameters, such as the mass ratio, inclinations, temperatures of primary and secondary components, modified surface potential, orbital period, and other physical properties of the binary stars. The process continues until a good fit is achieved, that means the residuals are minimized. DC2015 typically begins with the output from LC2015 and differential correction works by adjusting the model parameters in small steps to minimize the difference between the observed and calculated light curves. This process is iterative, where the model parameters are slightly modified, and the light curve is recalculated at each step. The differential correction process is stopped when the residuals stabilize or when the specified convergence criteria are met. The final output includes the refined values for the stellar parameters, with uncertainties or errors computed for each parameter, providing a robust and well-fitted model of the binary system.

In this research, the DC process enables the model to converge toward the most physically plausible solution by addressing errors in the initial parameters and iterating toward a more refined set of values. In conclusion, DC is a crucial step in binary star system modeling, as it allows for more precise parameter determination by iterating the model to reduce discrepancies with the observed light curves and it significantly improves the accuracy and reliability of the results.

## 7. REFERENCES

[1] Aliçavuş, F. K., Gümüş, D., Kırmızıtaş, Ö., Ekinci, Ö., Çavuş, S., Kaya, Y. T., & Aliçavuş, F. (2022). Candidate Eclipsing Binary Systems with a δ Scuti Star in Northern TESS Field. *Research in Astronomy and Astrophysics, 22(8), 085003*. doi:10.1088/1674-4527/ac71a4

[2] Darwish, M. S. (2020). Darwish, M. S., Abdelkawy, A. G., & Hamed, G. M. (2024). Light curve analysis and evolutionary status of four newly identified short-period eclipsing binaries. *Scientific Reports, 14(1), 3998*. doi:10.1038/s41598-024-54289-1

[3] Güzel, O. & Özdarcan, O. (2020). PyWD2015 -- A new GUI for the Wilson-Devinney code. *Contributions of the Astronomical Observatory Skalnaté Pleso, vol. 50, no. 2*, 535-538. doi:10.31577/caosp.2020.50.2.535

[4] Handberg, R., & Lund, M. N. (2014). Automated preparation of Kepler time series of planet hosts for asteroseismic analysis. *Monthly Notices of the Royal Astronomical Society, 445(3)*, 2698-2709. doi:10.1093/mnras/stu1823

[5] Kahraman Aliçavuş, F., Çoban, Ç. G., Çelik, E., Dogan, D. S., Ekinci, O., & Aliçavuş, F. (2023). Discovery of delta Scuti variables in eclipsing binary systems II. Southern TESS field search. *Monthly Notices of the Royal Astronomical Society, 524(1)*, 619-630. doi:10.1093/mnras/stad1898

[6] Kahraman Aliçavuş, F., Soydugan, E., Smalley, B., & Kubát, J. (2017). Eclipsing binary stars with a δ Scuti component. *Monthly Notices of the Royal Astronomical Society, 470(1)*, 915-931. doi:10.1093/mnras/stx1241

[7] Kallrath, J. (2022). Fifty Years of Eclipsing Binary Analysis with the Wilson–Devinney Model. *Galaxies, 10(1), 17*. doi:10.3390/galaxies10010017

[8] Liakos, A., & Niarchos, P. (2016). Catalogue and properties of δ Scuti stars in binaries. *Monthly Notices of the Royal Astronomical Society 465*, 1181-1200. doi:10.1093/mnras/stw2756

[9] Liakos, A., & Niarchos, P. (2020). Asteroseismic Analysis of δ Scuti Components of Binary Systems: The Case of KIC 8504570. *Galaxies, 8(4), 75*. doi:10.48550/arXiv.2010.10187

[10] Maceroni, C., Lehmann, H., Da Silva, R., Montalbán, J., Lee, C. U., Ak, H., Deshpande, R., Yakut, K.,Debosscher, J., Guo, Z., Kim, S.-L., Lee, J.W., & Southworth, J. (2014). KIC 3858884: a hybrid δ Scuti pulsator in a highly eccentric eclipsing binary. *Astronomy & Astrophysics, 563, A59*. doi:10.1051/0004-6361/201322871

[11] Soydugan, E., & Kacar, Y. (2013). Binarity and Pulsation in Algol-type Binary System SX Draconis. *The Astronomical Journal, 145(4), 87*, 8. doi:10.1088/0004-6256/145/4/87

[12] Stassun, K. G. (2019). The revised TESS input catalog and candidate target list. *The Astronomical Journal, 158(4), 138*, 21. doi:10.3847/1538-3881/ab3467

[13] Van Hamme, W. (1993). New limb-darkening coefficients for modeling binary star light curves. *Astronomical Journal (ISSN 0004-6256), vol. 106, no. 5*, 2096-2117.